\documentclass[aps,pre,twocolumn,amsmath,amssymb,superscriptaddress,groupedaddress]{revtex4-1}
\usepackage{bm} 
\usepackage{color}
\usepackage{amsmath}
\usepackage{amssymb}
\usepackage{mathrsfs}
\usepackage{graphicx} 
\usepackage{lastpage}
\usepackage{epstopdf}
\usepackage[normalem]{ulem} 
\usepackage[titletoc,title]{appendix}

\newcommand{\bcalF}{\mbox{\boldmath$\cal F$}}
\newcommand{\bcalG}{\mbox{\boldmath$\cal G$}}
\newcommand{\bCh}{\mathrm {\bf Ch}}

\begin{document}

\preprint{APS/123-QED}

\title{The role of symmetry in driven propulsion at low Reynolds number}

\author{Johannes Sachs$^{1,2}$} \author{Konstantin I. Morozov$^3$} \author{Oded Kenneth$^4$} \author{Tian Qiu$^{1}$} \author{Nico Segreto$^2$} \author{Peer Fischer$^{1,2}$}\email{fischer@is.mpg.de}
 \author{Alexander M. Leshansky$^3$}\email{lisha@technion.ac.il}

\affiliation{$^1$Max Planck Institute for Intelligent Systems, Heisenbergstra{\ss}e 3 70569 Stuttgart, Germany\\
$^2$Institute of Physical Chemistry, University of Stuttgart, Pfaffenwaldring 55, 70569 Stuttgart, Germany\\
$^3$Department of Chemical Engineering, Technion -- IIT, Haifa, 32000, Israel\\
$^4$Department of Physics, Technion -- IIT, Haifa, 32000, Israel}

\date{\today}

\begin{abstract}

We theoretically and experimentally investigate low-Reynolds-number propulsion of geometrically achiral planar objects that possess a dipole moment and that are driven by a rotating magnetic field. Symmetry considerations (involving parity, $\widehat{P}$, and charge conjugation, $\widehat{C}$) establish correspondence between propulsive states depending on orientation of the dipolar moment.
Although basic symmetry arguments do not forbid individual symmetric objects to efficiently propel due to spontaneous symmetry breaking, they suggest that the average ensemble velocity vanishes. Some additional arguments show, however, that highly symmetrical ($\widehat{P}$-even) objects exhibit no net propulsion while individual less symmetrical ($\widehat{C}\widehat{P}$-even) propellers do propel.
Particular magnetization orientation, rendering the shape $\widehat{C}\widehat{P}$-odd, yields unidirectional motion typically associated with chiral structures, such as helices. If instead of a structure with a permanent dipole we consider a polarizable object, some of the arguments have to be modified. For instance, we demonstrate a truly achiral ($\widehat{P}$- and $\widehat{C}\widehat{P}$-even) planar shape with an induced electric dipole that can propel by electro-rotation. We thereby show that chirality is not essential for propulsion due to rotation-translation coupling at low Reynolds number.
\end{abstract}

\maketitle

\section{Introduction}
Bacteria employ rotation-translation coupling when they spin their helical flagella in order to swim through fluids at low Reynolds (Re) numbers \cite{Turner}.
It is also possible to rotate an artificial magnetic corkscrew to generate propulsion \cite{GF,Nelson}. In both cases the rotation-translation coupling arises because of the symmetry-breaking due to the chiral shape. Purcell's famous remark ``Turn anything - if it isn't perfectly symmetrical, you'll swim" \cite{Purcell} raises the question
if a shape needs to be chiral to propel when it is spun at low Reynolds number. This issue is, of course, only relevant for swimming in unbounded liquid, as boundaries
can provide such coupling even for highly symmetrical driven objects, e.g., isotropic spherical ``microrollers"  \cite{chaikin17} and axisymmetric peanut-shaped colloids \cite{peanut18} exhibit net propulsion when rotated by an external field in the vicinity of a solid surface.

For a long time the geometric chirality of the object was taken for granted as a necessary condition for driven propulsion in rotating magnetic \cite{GF,Nelson,Schmidt,jacs13,acsnano14} and electric fields \cite{zeldovich,MD16} and for
cross-stream migration and separation in shear flows \cite{shear1,shear2,shear3} and for Brownian swimmers \cite{Lowen1}.
Recent experiments with random-shaped magnetic aggregates \cite{Vach1,Vach2}, as well as clusters made of just three magnetized beads \cite{Cheang} demonstrated propulsion in a uniform rotating magnetic field. The latter example is particularly intriguing, as it is  argued that seemingly ``achiral" planar objects are propulsive. Ref.~\cite{prf17} offered an explicit solution for driven rotation and propulsion of an arbitrarily shaped and magnetized object, establishing the dependence of the propulsion velocity on geometry and orientation of a permanent magnetic dipole moment. It had confirmed that geometric chirality is not required for driven propulsion and proposed that an achiral shape can break symmetry due to nontrivial orientation of its magnetic dipole. The most recent study \cite{scirob} showed that in agreement with the experiments \cite{Vach1,Vach2} the fastest random aggregates have a shape of an arc, i.e., having the same symmetry as the three-bead cluster of \cite{Cheang}.

In the present paper we address symmetry requirements for magnetized objects to propel when acted upon by an external torque and
show that chirality alone does not predict if an individual structure propels. Instead it only predicts whether a net propulsion persists after averaging over random initial orientations in a large ensemble of such structures. We apply these arguments to simple magnetic V-shapes with a permanent magnetic dipole moment affixed to it, and then demonstrate them experimentally. For polarizable objects with an \emph{induced} (rather than a permanent dipole) some of the symmetry arguments have to be revisited. We illustrate this experimentally considering electro-rotation of a planar V-structure with induced electrical dipole.

This research may have important practical implications, as  planar high-symmetry micro- and nanostructures are generally easier to fabricate than complex 3D low-symmetry shapes. The paper is organized as follows: Sec.~\ref{sec:rotmag} briefly outlines the governing equations of motion and their explicit solution. Sec.~\ref{sec:sym} discusses the relevant symmetries of the solution and in Sec.~\ref{sec:planar} the respective consequences of the symmetries for the dynamics of a magnetized V-shape with a permanent dipole are theoretically investigated. Sec.~\ref{sec:electro} addresses the close analogy between electric and magnetic propellers. Finally, experiments in support of the theory are provided in Sec.~\ref{sec:exp}.

\section{Propulsion driven by a rotating magnetic field \label{sec:rotmag}}
We consider the propulsion of an arbitrary magnetic object driven by an external uniform magnetic field
${\bm H}=H(\hat{\bm x}\cos \omega t+\hat{\bm y}\sin \omega t)$.
At low Re an object's velocity is linearly dependent on the forces and torques exerted on it.
If no external force is applied, then only the external magnetic torque, $\bm{L}=\bm m\times \bm H$, can be responsible for the actuation of the object:
\begin{equation}
\bm{U}=\bcalG \cdot \bm{L}\,,\qquad \bm{\mathit{\Omega}}={\bcalF}\cdot \bm{L}\,, \label{eq:1}
\end{equation}
where $\bm{U}$ and $\bm{\mathit{\Omega}}$ are the object's translation and rotation velocities, ${\bcalG}$ the rotation-translation coupling
mobility (pseudo-)tensor and $\bcalF$ is the rotational mobility tensor.
Typical dynamics in a \emph{rotating} magnetic field (provided that the rotation frequency $\omega$ is not too high)
exhibit, after a short transient period, a solution where the the object is turning in-sync with the
actuating field., i.e., rotating about the $z$-axis with angular velocity $\bm{\mathit{\Omega}}={\omega}\hat{\bm z}$.
It was found that there can be up to two stable synchronous solutions of Eqs.~(\ref{eq:1}) \cite{prf17}.

Expressing the magnetic torque, $\bm{L}$, using the second Eq. in  (\ref{eq:1}) and substituting it into the the first Eq. in (\ref{eq:1}), the translational velocity $\bm{U}$ can be readily found as ${\bm U}={\bcalG}\cdot{\bcalF}^{-1}\cdot \bm{\mathit{\Omega}}$. By symmetry the average velocity in an in-sync solution is along the $z$-axis.
Taking a scalar product on both sides of this equation with $\bm{\mathit{\Omega}}=\omega \hat{\bm z}$ we readily obtain it in a compact covariant form as \cite{prf17,scirob}
\begin{equation}
\frac{U_z}{\omega\ell}=\widehat{\bm{\mathit{\Omega}}}\cdot\bCh
\cdot\widehat{\bm{\mathit{\Omega}}}\;,\label{eq:Uz}
\end{equation}
where $\bCh$ is a dimensionless \emph{chirality matrix} given by the symmetric part of $\frac{1}{\ell}\, \bcalG \cdot \bcalF^{-1}$ with $\ell$
being the characteristic length and $\widehat{\bm{\mathit{\Omega}}}=\bm{\mathit{\Omega}}/\omega=\hat{\bm z}$ the normalized (unit) angular velocity.
It  is most convenient to write the RHS of (\ref{eq:Uz}) in the body frame spanned by a triad of body frame unit vectors,
$\left\{{\bm e}_1, {\bm e}_2, {\bm e}_3\right\}$
affixed to the rotating body.  These are defined to be the principal rotation axes, i.e. the
eigenvectors of ${\bcalF}$.  The lab coordinates unit vectors $\{\hat{\bm x},\hat{\bm y},\hat{\bm z}\}$ are related to
the body frame axes by a rotation matrix ${\cal R}(\varphi,\theta,\psi)$ parameterized through the three Euler angles $\varphi,\,\theta$ and $\psi$. Thus, in the body frame $\bCh$ is fixed and $\widehat{\bm{\mathit{\Omega}}}$ expressed via the Euler angles. Note that $\bCh$ (in contrast to $\bcalG$) is independent of the choice of coordinate origin. Under rotation of the coordinate frame it transforms
as a (symmetric) pseudo-tensor.

The Euler angles $\psi,\theta,\varphi$ (and the rotation matrix ${\cal R}$) are determined by solving the second
Eq. in  (\ref{eq:1}). For synchronous solutions $\dot{\psi}=\dot{\theta}=0$, $\dot{\varphi}=\omega$ this turns into a system of algebraic equations for  constant values of $\psi, \theta$ and $\tilde{\varphi}=\varphi-\omega t$ (see Appendix~\ref{A}).

\section{Chirality and symmetries of solutions \label{sec:sym}}

In general, the symmetries of the object's shape determine the structure of its resistance/mobility tensors ${\bcalF},{\bcalG}$ \cite {HB}.
For externally driven objects it is not sufficient to consider only the shape of the object, but one also needs to include the transformation
property of its \emph{dipole moment} ${\bm m}$ \cite{Barron}. Any proper discussion of the symmetries of Eqs.~\ref{eq:1} and their solutions
must involve both a discussion of  the symmetries of the swimming object as well as those of the external magnetic field ${\bm H}$.

As the magnetized object is actuated by an externally applied magnetic field, the equations governing its evolution are invariant only under symmetries which preserve this field.
The magnetic field ${\bm H}=H(\hat{\bm x}\cos \omega t+\hat{\bm y}\sin \omega t)$ is invariant under three independent symmetries.
(i)  $\widehat{P}$ (ii) $\widehat{C}\widehat{R}_z$ (iii) $\widehat{T}\widehat{R}_y$.
Here $\widehat{P},\widehat{C},\widehat{T}$ denote parity, charge conjugation and time reversal, respectively, while $\widehat{R}_x,\widehat{R}_y,\widehat{R}_z$ denote
rotation by $\pi$ around the lab coordinate axes. Below we shall also use the notation  $\widehat{R}_1,\widehat{R}_2,\widehat{R}_3$  to denote rotation by $\pi$ around the body frame principal axes ${\bm e}_1, {\bm e}_2, {\bm e}_3$, respectively.

The last symmetry (iii) involving time reversal $\widehat{T}$ maps stable in-sync solutions into unstable solutions.
As we are only interested in stable solutions, this symmetry will be irrelevant to our considerations and we shall not discuss it further \cite{time_rev}.
We note that in our present context of magnetically driven propulsion charge conjugation $\widehat{C}$ corresponds to a reversal of the dipole moment,
while parity transformation (or point reflection) is described by the operator
$\widehat{P}$,
\begin{equation}
\widehat{P}(x,y,z) \rightarrow (-x,-y,-z)  \,. \label{eq:2}
\end{equation}
For future reference we note that the relevant quantities governing the dynamics
exhibit the following transformation properties under $\widehat{P}$ and $\widehat{C}$:
\begin{eqnarray}
(\bcalF,\bcalG,\bm{m},\bm{H}) &\overset{\widehat{P}}{\mapsto}& (+\bcalF,-\bcalG,+\bm{m},+\bm{H})\,,  \\
(\bcalF,\bcalG,\bm{m},\bm{H}) &\overset{\widehat{C}}{\mapsto}& (+\bcalF,+\bcalG,-\bm{m},-\bm{H})\,.
\end{eqnarray}

Since the actuating magnetic field $\bm H$ is invariant under parity, any solution of Eqs.~\ref{eq:1}
for a magnetic propeller must then be mapped to a valid solution under the action of $\widehat{P}$.
As parity $\widehat{P}$ is not a proper rotation, it cannot be practically implemented on a physical 3D object. An object is called \emph{achiral} ($\widehat{P}$-even) if there exists a proper rotation $\widehat{R}$ whose action on it is equivalent to the action of parity, (and \emph{chiral} otherwise).
In such a case the action of parity is equivalent to a proper rotation.
Applying this constant rotation $\widehat{R}$ then maps a solution to another solution of Eqs.~\ref{eq:1}.
Thus merely rotating an achiral object through  $\widehat{R}$ would result in a parity-dual solution having the reverse propulsion velocity.

We note that in terms of ${\cal R}(\varphi,\theta,\psi)$ this map is
 ${\cal R}\mapsto{\cal R}\widehat{R}$ (as $\widehat{R}$ is defined in the body frame).
The case in which $\widehat{R}=I$ correspond to a parity-symmetric object which clearly cannot swim as it has vanishing coupling matrix $\bcalG=0$.
This trivial case is of no real interest. In any nontrivial (achiral) case $\widehat{R}\neq I$ implies ${\cal R}\neq{\cal R}\widehat{R}$, showing that the two solutions are truly distinct.
Since there are no more than two stable solutions, it follows that these are all the solutions in this case.

\begin{figure*}
\centering
\includegraphics[scale=1]{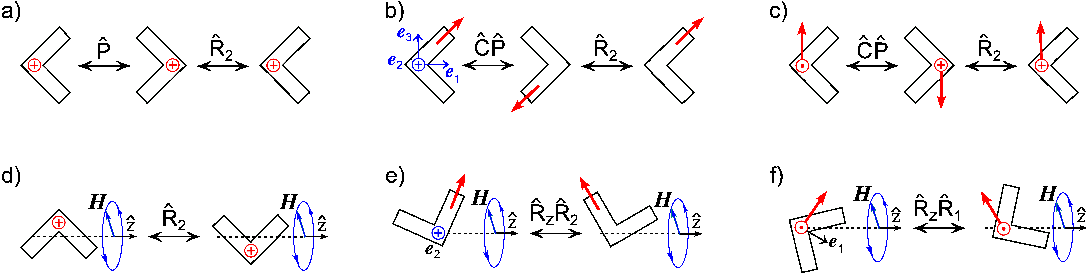}
\caption{\label{fig:parity_scheme} Schematic diagrams illustrating symmetries of a planar V-shaped object carrying a magnetic
dipole moment (red) of different orientations with respect to parity, $\widehat{P}$, and
charge conjugation, $\widehat{C}$. (a) $\widehat{P}$- (and $\widehat{C}\widehat{P}$-) even object;
(b) $\widehat{C}\widehat{P}$-even (and $\widehat{P}$-odd) object;
(c) $\widehat{C}\widehat{P}$- (and $\widehat{P}$-) odd object.
Principal axes of rotation $\{\bm{e}_1, \bm{e}_2, \bm{e}_3\}$ are shown in (b).
(d-f) diagrams illustrating the pairs of solutions corresponding to (a-c).}
\end{figure*}

A rotating magnetic field, however, possesses also the extra symmetry (ii): $\widehat{C}\widehat{R}_z$.
One can easily check that combining a rotation $\widehat{R}_z$ by $\pi$ around the $z$-axis,
with charge conjugation $\widehat{C}$ maps $\bm H$ to itself and is thus a symmetry
\cite{phase}.
It follows that $\widehat{R}_z\widehat{C}\widehat{P}$ is a symmetry too.
In analogy to the usual notion of chirality, we shall call an object $\widehat{C}$-even or
$\widehat{C}\widehat{P}$-even if there exist a rotation
$\widehat{R}$ such that the object is symmetric under $\widehat{R}^{-1}\widehat{C}$ or
$\widehat{R}^{-1}\widehat{C}\widehat{P}$ respectively.

Being  $\widehat{C}\widehat{P}$-even implies that one can substitute the action of
$\widehat{C}\widehat{P}$ by the action of a proper rotation $\widehat{R}$.
Thus the symmetry $\widehat{R}_z\widehat{C}\widehat{P}\sim\widehat{R}_z\widehat{R}$
shows that changing the orientation of a $\widehat{C}\widehat{P}$-even object
(by applying $\widehat{R}_z\widehat{R}$-rotation)
maps a swimming solution to a swimming solution.
As $\widehat{C}\widehat{P}$ reverses velocities we obtain  a pair of dual solutions having opposite propulsion velocities.
In terms of ${\cal R}(\varphi,\theta,\psi)$ this map is ${\cal R}\mapsto\widehat{R}_z{\cal R}\widehat{R}$  (as $\widehat{R}$ is defined in the body frame and $\widehat{R}_z$ in the lab frame).
The two solutions are in general distinct. Indeed they can coincide only if the two rotations
$\widehat{R}_z,\widehat{R}$ cancel each other which can happen only if the (body frame)
rotation axis of $\widehat{R}$ coincides with the lab $z$-axis.

For a $\widehat{C}$-even object the same arguments lead to the existence of pairs of solutions having identical velocities.
If the solutions are distinct (as must happen whenever the
rotation axis of $\widehat{R}$ does not coincides with the lab $z$-axis) than these are necessarily all the stable solutions.

It may also be noted that the symmetry (ii) under $\widehat{C}\widehat{R}_z$ (and hence also
under $\widehat{C}\widehat{P}\widehat{R}_z$) is special to the case of a rotating field
${\bm H}=H(\hat{\bm x}\cos \omega t+\hat{\bm y}\sin \omega t)$. Adding, for example, a constant field component along the $z$-axis
is enough to break it and some of the results we derive below will not apply in this case.
For the experiment described in this paper this symmetry is very relevant.

\section{Dynamics of planar V-shaped propellers \label{sec:planar}}

We apply the above symmetry arguments to planar V-shaped objects, schematically depicted in Fig.~\ref{fig:parity_scheme} and then test these predictions experimentally in Sec.~\ref{sec:exp}. Note that the V-shape (ignoring its dipole moment) is a highly symmetrical object with two mutually perpendicular symmetry planes.
In the frame of \emph{principal rotation axes} aligned with eigenvectors of $\bcalF$
[see Fig.~\ref{fig:parity_scheme}(b)]
it therefore has only  two nontrivial  component of $\bcalG$ ($\mathcal{G}_{23}$ and $\mathcal{G}_{32}$)
\cite{prf17}.

Consider first a V-shaped structure with a magnetic moment $\bm m$ oriented perpendicular to the plane of the object [see Fig.~\ref{fig:parity_scheme}(a)].
This object is $\widehat{P}$-even as well as $\widehat{C}\widehat{P}$-even.
Denoting by $\widehat{R}_i$ rotation by $\pi$ around the $i$th principal rotation axis,
one can see that the object is mapped onto itself under $\widehat{R}_2\widehat{P}$, $\widehat{R}_1\widehat{C}$ and $\widehat{R}_3\widehat{C}\widehat{P}$.
(Notice, however, that asymmetric V-shape with unequal arms is only $\widehat{P}$-even.) When the magnetic moment lies in plane of the V-shape, e.g., directed along one of the arms as shown in Fig.~\ref{fig:parity_scheme}(b), the object is $\widehat{P}$-odd, but $\widehat{C}\widehat{P}$-even as
$\widehat{R}_2\widehat{C}\widehat{P}$ maps it onto itself. The object in Fig.~\ref{fig:parity_scheme}(c) with an off-plane orientation of the dipole,
on the other hand, is odd under both $\widehat{P}$ and $\widehat{C}\widehat{P}$;
being invariant under $\widehat{R}_1\widehat{C}$ it is $\widehat{C}$-even.

Let us focus on the details of the propulsion.
Fig.~\ref{fig:parity_scheme}(e) illustrates the existence of two propulsive states related by the
symmetry of a $\widehat{C}\widehat{P}$-even object. We assume arbitrary orientation of the V-shape with respect to the field so that its rotation
could be accompanied by precession (or \emph{wobbling}).
As $\widehat{C}\widehat{P}\sim \widehat{R}_2$, the $\widehat{R}_z\widehat{C}\widehat{P}$-symmetry
implies that $\widehat{R}_z\widehat{R}_2$-rotation leads to another solution.
Since $\widehat{C}\widehat{P}$ inverts linear velocities, the two solutions have opposite propulsion velocities.
As we shall demonstrate below, an individual $\widehat{C}\widehat{P}$-even objects can propel quite efficiently. The propulsion direction, $+z$- or $-z$,
is controlled by the initial orientation which serves as to `spontaneously break the symmetry'.  Thus a large collection of such propellers having random initial orientations would at most exhibit symmetric spreading with zero ensemble average velocity, as if it was a \emph{racemic mixture} having an equal number of structures with opposite handedness.
Notice that the symmetry arguments concerning the V-shaped object in Fig.~\ref{fig:parity_scheme}(b) apply even if its two arms are unequal.


Similar arguments can be applied to the highly symmetric $\widehat{P}$-even object with a permanent magnetic dipole
${\bm m}$ as in Fig.~\ref{fig:parity_scheme}a. A stronger result, however, can be obtained in this case combining symmetry and geometric arguments.

Although the shape of an arbitrary object can be infinitely complicated, in the low Reynolds-number regime we are guaranteed that
its propulsion in the magnetic field is fully described in terms of a limited number of variables. Namely a (positive symmetric) tensor $\bcalF$ a pseudo-tensor $\bcalG$ (which w.l.o.g is also symmetric) and the pseudo vector $\bm{m}$ (in the case of a permanently magnetized object). Moreover the rotational equations involve only $\bcalF$ and $\bm{m}$.
These facts put further restrictions on the possible behavior of chiral objects.
Notice that if instead of an object with permanent dipole we have a \emph{polarizable} (i.e., superparamagnetic) object, then the analysis below has to be modified.

Let us first show that any $\widehat{P}$-even magnetized object either has ${\bcalG}=0$
(implying it cannot propel upon rotation) or $\bm{m}\|\bm{e_i}$. To see this note that invariance under parity implies the existence of a rotation $\widehat{R}$ satisfying
\[
\widehat{R}\,\bcalF\,\widehat{R}^{-1}=\bcalF\,,\quad \widehat{R}\,\bcalG\,\widehat{R}^{-1}=-\bcalG\,,\quad \widehat{R}\,\bm{m}=\bm{m}\,.
\]
The first relation implies that $\widehat{R}$ is diagonal in the frame of principal axes $\left\{{\bm e}_1, {\bm e}_2, {\bm e}_3\right\}$.
If $\widehat{R}=I$ then from the second relation follows ${\bcalG}=0$.
Otherwise $\widehat{R}$ must be a rotation by $\pi$ around one of the principal axes $\bm{e}_i$ and the
third relation then demands $\bm{m}$ to be along this axis.
Note, moreover, that the relation $\widehat{R}\bcalG\widehat{R}^{-1}=-\bcalG$
with $\widehat{R}=\widehat{R}_i$ a $\pi$-rotation around $\bm{e}_i\|\bm{m}$ limits further the form of $\bcalG$.
If, e.g., $i=3$ it follows that only the elements ${\mathcal G}_{13}$, ${\mathcal G}_{31}$, ${\mathcal G}_{23}$ ${\mathcal G}_{32}$ can be nonzero.

Next we demonstrate that an object having  $\bm{m}$ along a principal axis, [e.g., along $\bm e_2$, as in Fig.~\ref{fig:parity_scheme}(b
)], will necessarily rotate around some other of its principal axes. To see this note that as $\bm{L}=\bm{m}\times\bm{H}$ is orthogonal to $\bm{e}_2\|\bm{m}$ so must also be $\bm{\mathit{\Omega}}=\bcalF \cdot \bm{L}$. As both $\bm{L}$ and $\bm{\mathit{\Omega}}$ are also orthogonal to $\bm{H}$ we conclude that they are parallel (unless $\bm{L}=\bm{m}\times\bm{H}=0$).
The relation $\bm{\mathit{\Omega}}=\bcalF\cdot\bm{L}$ then implies that $\bm{\mathit{\Omega}}$ is an eigenvector of $\bcalF$ and hence coincides with one of the principal axes.

We further notice that an object undergoing planar precession-free rotation (\emph{tumbling}) around a principal axis $\bm{e}_i$
must also have the torque $\bm{L}=\bcalF^{-1}\cdot \bm{\mathit{\Omega}}$ along this axis. Thus it implies that ${\bm e}_i$, $\bm{\mathit{\Omega}}$ and ${\bm L}$ are all parallel to $\hat{\bm z}$. Recalling that the net propulsion is also along the $z$-axis, we find that $U_z\propto\mathcal{G}_{ii}$ (no summation) vanishes for any geometrically achiral shape, e.g., V-shaped object. In particular, neither of the two tumbling solutions shown in Fig.~\ref{fig:parity_scheme}(d) and related by $\widehat{P}\sim \widehat{R}_2$ yield any net propulsion. Notice that for a symmetric V-shape, the solutions in Fig.~\ref{fig:parity_scheme}(d) are invariant under $\widehat{R}_z\widehat{C}\widehat{P}$ which also proves that $\bm{U}=0$. However, our arguments are more general (they apply, e.g., to $\widehat{P}$-even V-shape with unequal arms.)

Combining the above arguments it follows that $\widehat{P}$-even magnetic swimmers with permanent dipole moment $\bm{m}$ cannot propel. Furthermore, any geometrically achiral shape (such as V-shape), rotating around a principal axis in-sync with the actuating field will exhibit no net propulsion \emph{regardless of its symmetry}, The net propulsion, in general, requires precession or \emph{wobbling} of the object.

The finding that magnetized $\widehat{P}$-even objects are unable to propel should be compared with the weaker result we had for a magnetized $\widehat{C}\widehat{P}$-even propellers [see Fig.~\ref{fig:parity_scheme}(b),(e)] that can propel individually, while only the average velocity (over random initial orientations) needs to vanish.
Notice also that the above reasoning only holds for structures with a permanent dipole. As we shall demonstrate below
individual achiral ($\widehat{P}$-even) \emph{polarizable} objects can propel, as parity symmetry only guarantees vanishing of their ensemble average velocity.

Finally, the least symmetric ($\widehat{P}$- and $\widehat{C}\widehat{P}$-odd) objects
exhibit enantiomeric selection of the propulsion direction even when averaged over arbitrary initial orientation.
This is illustrated for the V-shape in Fig.~\ref{fig:parity_scheme}(c) with the magnetic moment, $\bm m$,
lying in the plane orthogonal to $\bm e_1$. The two solutions shown in Fig.~\ref{fig:parity_scheme}(f)
can be related by the symmetry $\widehat{R}_z\widehat{C}$. As $\widehat{C}\sim \widehat{R}_1$
the V-shape is $\widehat{C}$-even and applying
$\widehat{R}_z\widehat{R}_1$ yields another valid solution.
Since  $\widehat{C}$ (as opposed to $\widehat{P}$ and
$\widehat{C}\widehat{P}$) does not invert velocities, it guarantees that the two
solutions possess the same propulsion velocity $\bm U$.
Thus, the propulsion direction depends on the sense of rotation in exactly the same way as for left- or right-handed helices:
the original object in Fig.~\ref{fig:parity_scheme}(c) will translate along the field rotation $z$-axis under clockwise (CW) rotation,
while its $\widehat{C}\widehat{P}$-transformed enantiomer would propel in the opposite direction, regardless of the initial orientation.
Notice that a collection of less symmetric ($\widehat{C}$-odd) propellers, e.g., with unequal arms would still exhibit nonzero ensemble average velocity.

\section{Symmetries of electro-rotation solutions \label{sec:electro}}

Since an electric field $\bm{E}$ acts on an electric dipole $\bm{p}$ in exactly the same way that magnetic
field $\bm{H}$ acts on a magnetic dipole $\bm{m}$, one could imagine an electric propeller which is a complete analog of the magnetic one described above.
This case differs from magnetic case described earlier in a number of aspects.

We first notice that the transformation properties of $\bm{E}$ and $\bm{p}$ under $\widehat{P}$ ($\widehat{C}\widehat{P}$, respectively) are exactly the same
as the transformation properties of $\bm{H}$ and $\bm{m}$ under $\widehat{C}\widehat{P}$ ($\widehat{P}$, respectively).
For a hypothetical object with permanent electric dipole actuated by a rotating electric field, one could apply the same arguments
we had in the magnetic case using $\widehat{C}\widehat{P}$ symmetry instead of $\widehat{P}$ (and vice versa).
The conclusion would then also require merely a change of terminology $\widehat{C}\widehat{P}\leftrightarrow\widehat{P}$ everywhere.

Having a propeller with a permanent electric dipole is, however, not practically feasible and below (see Sec.~\ref{sec:exp}) we present experiments involving polarizable structures with an \emph{induced} electric dipole. Yet, some of the above arguments (that use the specific form of the governing equations) do not hold for polarizable (either electric or magnetic) propellers.

\section{Experimental results \label{sec:exp}}
\begin{figure}
\includegraphics[scale=0.8]{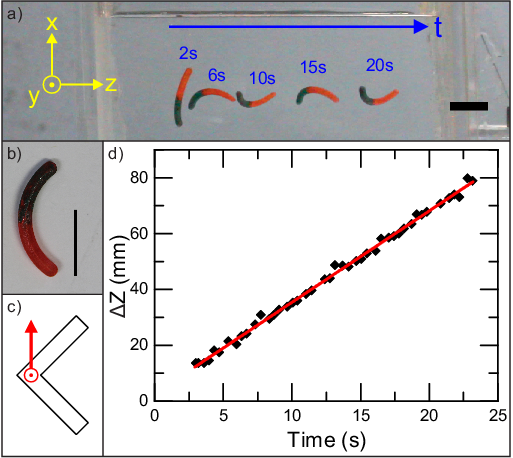}
\caption{\label{fig:timeevo}  (a) Snapshots of the magnetically driven arc.
The magnetic field rotates clockwise in the $xy$-plane with a frequency of $1.5$~Hz
resulting in the arc rotating and propelling along the $z$-axis.
(b) Image of the arc; (c) a corresponding  symmetry diagram; (d) arc displacement along the $z$-axis vs. time
showing a constant speed of $U_{z}\approx 3.3$~mm/s (the scale bar length in (a) and (b) is 1~cm).}
\end{figure}

We now experimentally demonstrate the different propulsion gaits associated with different symmetries. We perform low-Reynolds-number experiments with cm-sized magnetic structures immersed in glycerol, as well as $\mu$m-sized structures that possess either a magnetic or an induced electric dipole moment suspended in water. The larger objects allow for precise positioning and alignment of the magnetic moment.
An arc-shaped structure with cross-sectional radius $a=1$~mm was 3D-printed [see Fig.~\ref{fig:timeevo}(b)]. It had a cubic compartment into which a small 1~mm$^3$  NdFeB (N45) ferromagnet was glued. The orientation of the magnet with respect to the object was therefore fixed and prescribed. The arc was placed
in a cuvette filled with glycerol ($\eta\approx 1,000$~cP). The high viscosity prevented sedimentation and ensured
that $\mathrm{Re\approx 1}$. A pair of two disk-shaped iron-based permanent magnets generated a homogeneous magnetic field of $300$~G throughout the volume of the cuvette. The magnets were mounted and mechanically rotated in the $xy$-plane around the cuvette. The driven motion of the arc was recorded and analyzed. Results for a right-handed arc with an off-plane
magnetization [as in Fig.~\ref{fig:parity_scheme}(c)] actuated by a rotating field at frequency of 1.5~Hz are shown in Fig.~\ref{fig:timeevo}. In Fig.~\ref{fig:timeevo}(a)
the position and orientation of the arc is depicted at different times; Fig.~\ref{fig:timeevo}(d) shows the corresponding displacement of the arc's centerpoint along the
$z$-axis of the field rotation (see Video~\#1 \cite{SM}). The arc turns in-sync with the field and propels along the $z$-axis, as expected for CW rotation.

\begin{figure}[tb]
\includegraphics[scale=0.80]{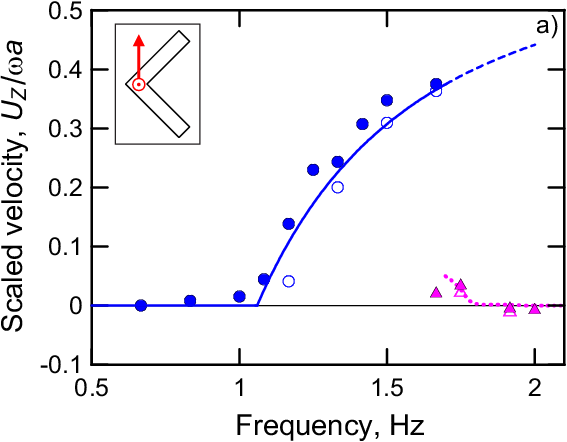} \\
\includegraphics[scale=0.80]{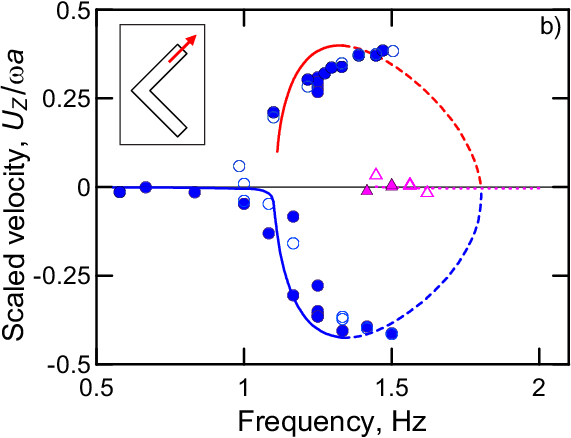}
\caption{\label{fig:V_vs_freq} Scaled propulsion velocity $U_z/\omega a$ of the magnetic arc vs. the actuation frequency $\omega/2\pi$: (a) $\widehat{C}\widehat{P}$-odd off-plane magnetized propeller; (b) $\widehat{C}\widehat{P}$-even in-plane magnetized propeller. The insets show the corresponding symmetries. Circles stand for in-sync actuation and triangles for asynchronous tumbling. Filled and empty symbols correspond to clockwise and counter-clockwise rotation of the magnetic field, respectively. Lines are the theoretical predictions corresponding to stable in-sync actuation with infinite (solid) and finite (dashed) basin of attraction and to aperiodic tumbling (dotted).}
\end{figure}

In Fig.~\ref{fig:V_vs_freq}(a) the scaled propulsion velocity $U_{z}/\omega a$ of an off-plane
magnetized ($\widehat{C}\widehat{P}$-odd) arc is depicted vs. the actuation frequency
$\nu=\omega/2\pi$.
At low frequencies $\nu\lesssim 1.1$~Hz the arc tumbles without any noticeable translation. Above
the tumbling-to-wobbling transition frequency,
$\nu_{\mathrm {t\mbox{-}w}}\sim 1.1$~Hz, it starts to precess and propel along the $z$-axis.
The fact that propulsion occurs only at $\nu>\nu_{\mathrm {t\mbox{-}w}}$, demonstrates its dependence on \emph{dynamics} rather than just symmetry.
In Figs.~\ref{fig:parity_scheme}(e, f) the non-propulsive tumbling regime corresponds to vanishing of the angle
between $\bm{e}_1$ and $\hat{\bm z}$, that is precession-free rotation around $\hat{\bm z}$. The regime selection at a given actuation frequency, i.e., non-propulsive tumbling vs. propulsive wobbling, depends on whether it is energetically more favorable for the angle between $\bm{e}_1$ and $\hat{\bm z}$ to vanish or to have nonzero value. Such energy arguments were provided in \cite{ML14a} assuming cylinder-like rotational anisotropy of the driven object.
The direction of translation is controlled by the rotation sense of the applied magnetic field,
thus the velocity of the structure in Fig.~\ref{fig:V_vs_freq}(a) is always positive.
The velocity increases quasi-linearly with frequency,
$U_z \sim \nu a\, (1-\nu_{\mathrm{t\mbox{-}w}}^2/\nu^2)$, similarly to a magnetic
helix \cite{ML14a} up to $\nu \sim1.6$~Hz
in excellent agreement with the theory (see Appendices~\ref{B} and \ref{C}).
For frequencies $\nu \gtrsim1.6$~Hz the arc can no longer turn in-sync with the
external field and exhibits asynchronous twirling accompanied by a
negligible net propulsion \cite{ML14a}.

Note that while a combined rescaling of both $\omega$ and ${\bm H}$ is expected to lead to a self-similar solution,
the results presented here as a function of $\omega$ correspond to fixed value of $H$
which, e.g., sets the scale of $\nu_{\mathrm {t\mbox{-}w}}\propto H m \mathcal{F}_\perp$, where $\mathcal{F}_\perp$ is the transverse rotational mobility (see Appendix~\ref{B}).
Thus no simple linear relation between $U$ and $\omega$ is expected.

In Fig.~\ref{fig:V_vs_freq}(b) the velocity-frequency dependence is shown for an arc with a magnetic moment oriented
along one of the arms. This ($\widehat{C}\widehat{P}$-even) structure can
spontaneously break symmetry and exhibit translation when actuated
at $\nu>\nu_{\mathrm {t\mbox{-}w}}$.
Symmetry demands, however, that for every initial orientation of the structure propelling in one direction, there exists an orientation for which it will propel in the opposite direction as the symmetry transformed counterpart.
The experimental results showing a symmetric pitchfork bifurcation in Fig.~\ref{fig:V_vs_freq}(b) (symbols) agree very well with the theory (see Appendices~\ref{B} and \ref{C})
and the arc can propel in the $\pm z$ direction irrespective of the sense of magnetic field rotation.
It has also been confirmed experimentally (see Appendix~\ref{D} and Video~\#2 \cite{SM})
that upon reversal of the field rotation, the object maintains its propulsion direction.
Additional experiments with the achiral arc magnetized along the principal rotation axes
[e.g., Fig.~\ref{fig:parity_scheme}(a)] have been performed demonstrating no propulsion, as expected (see Appendix~\ref{E} and Video~\#3 \cite{SM}).
\begin{figure}[t]
\includegraphics[scale=0.80]{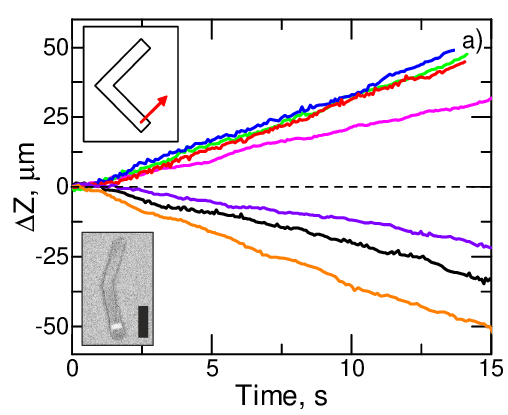}\\
\includegraphics[scale=0.80]{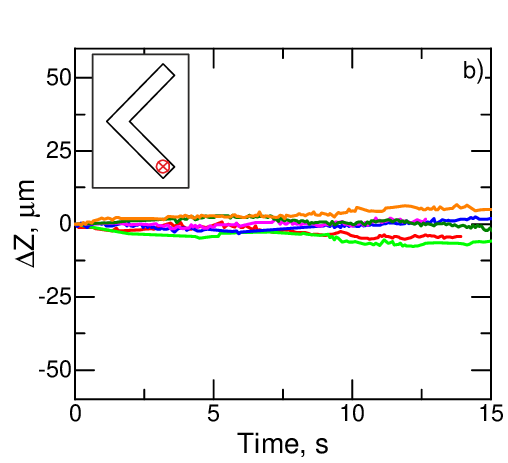}
\caption{Displacement $\Delta z$ of V-micropropellers with different magnetization and therefore different symmetries which are schematically depicted in the corresponding (top) insets for each case. (a) Individual $\widehat{C}\widehat{P}$-even and $\widehat{P}$-odd (chiral) structures move with constant velocity either in the $+z$ or the $-z$-direction; (b) $\widehat{C}\widehat{P}$- and $\widehat{P}$-even (achiral) structures show no net translation upon rotation. Bottom inset in (a) is a SEM image of the
micropropeller. The nickel segment (white) is clearly visible in the SiO$_2$ (grey) arm. Scale bar length is 1~$\mu$m. \label{fig:GLAD_arc} }
\end{figure}

To demonstrate applicability of the symmetry considerations to the microscale objects, we used a physical vapor deposition method, known as glancing angle deposition (GLAD)
to grow billions of magnetic microstructures on a wafer \cite{Robbie,GF}. V-shaped SiO$_2$ microstructures containing a nickel section were grown onto silica beads
(see SEM image in the inset in Fig.~\ref{fig:GLAD_arc}(a)). The growth direction is well-controlled during the GLAD process,
which enables us to orient the V-shaped structures before they are magnetized. The desired magnetization was obtained by placing the wafer with the structures in an
electromagnet (1.8~T) at a specific angle. Afterwards the V-shaped structures were removed from the wafer in an ultrasonic bath and dispersed in a solution
of 150~$\mu$M poly(vinylpyrrolidone). A custom 3-axis Helmholtz-coil setup was put up a microscope to generate a uniform magnetic field of $60$~G, rotating
at a frequency of $25$~Hz in the $xy$-plane. Slight variation in the shape and the direction of magnetization of the colloids are expected. Two microstructures
with symmetries shown in Fig.~\ref{fig:parity_scheme}(a) and Fig.~\ref{fig:parity_scheme}(b) were investigated and their translation along the $z$-axis of the field rotation was measured. It is also possible to control the trajectory of the micropropellers by switching the external field rotation plane between all three principal planes, including, e.g., displacement out-of-focus of the microscope (see Appendix~\ref{F} and Video~\#4 \cite{SM}).
In Fig.~\ref{fig:GLAD_arc}(a) the displacement $\Delta z$ of several $\widehat{C}\widehat{P}$-even structures is plotted vs. time. It can be seen that they move in opposite directions with approximately the same speed of $U_{z}\approx 2.7$~$\mu$m/s, which is about one body-length per second (see Video~\#5 \cite{SM}). On the other hand, as illustrated in  Fig.~\ref{fig:GLAD_arc}(b), net propulsion of the achiral V-micropropellers is negligible as expected. These experimental observations are thus in agreement with the symmetry arguments above.
\begin{figure}[tb]
\includegraphics[scale=0.85]{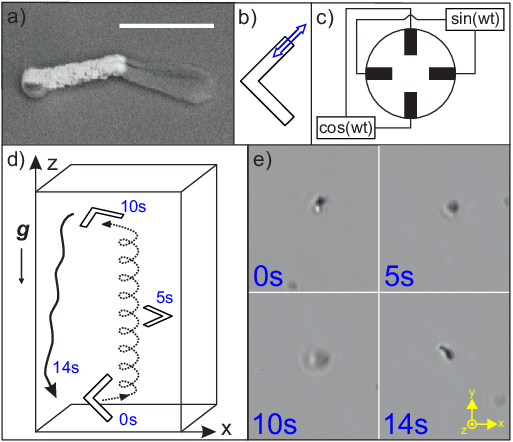}
\caption{\label{fig:electric_arc} (a) SEM image of polarizable SiO$_2$ V-micropropeller with a Au-coated (white) arm (scale bar length is 1~$\mu$m) and (b)  schematic drawing (double-headed arrow indicates the polarization easy-axis). (c) Sketch of the four-electrode setup generating a rotating electric field in the $xy$-plane. (d) Schematic of time evolution showing the displacement along the $z$-axis.
(e) First, the structure appears in focus ($0$~s), then moves out-of-focus ($5$~s and $10$~s) and after the electric field is turned off it sediments back into the focal plane due to gravity ($14$~s).}
\end{figure}

Finally, we demonstrate that propulsion can occur for structures of even higher symmetry. For that purpose we conducted experiment that involved polarizable structures with induced (rather than permanent) electric dipole. The GLAD technique was applied to grow V-shaped SiO$_2$-microstructures with one arm being coated with a thin layer of gold, as shown in in Fig.~\ref{fig:electric_arc}(a). Au is highly polarizable and should, therefore, give rise to an induced electric dipole moment $\bm{p}=\bm{\alpha}\cdot \bm{E}$, where $\alpha$ is a tensorial polarizability. Due to the anisotropy of the Au patch the V-microstructure [see Fig.~\ref{fig:electric_arc}(b)] possesses an anisotropic polarizability $\alpha_{\|}>\alpha_{\perp}$, which tends to align the Au-coated arm parallel to the external rotating $\bm{E}$-field.

After sonication in deionized water the suspension was placed in a four-electrode setup as schematically shown in Fig.~\ref{fig:electric_arc}(c) (see also Appendix~\ref{G}. Applying sinusoidal $\pi/2$ out-of-phase potentials at the electrodes at 500~kHz results in a rotating AC electric field $\bm E$ that exerts an electric torque on the V-shape, $\bm{L}=\bm{p}\times \bm{E}=(\alpha_{\|}-\alpha_{\perp})[\bm{n}\times \bm{E}](\bm{n}\cdot \bm{E})$, where $\bm n$ is a unit vector along the Au-arm, causing its electro-rotation \cite{Fan}. The actuating AC frequency of $500$~kHz is definitely beyond the step-out which is typically some tens of Hz, and a slow quasi-steady aperiodic rotation is taking place. General symmetry arguments should, however, hold regardless of the actuation regime. The resultant propulsion of the polarizable V-microstructure (against gravity) is illustrated in Figs.~\ref{fig:electric_arc}(d), (e) (see also Appendix~\ref{G} and Video~\#6, \cite{SM}) and it obviously resembles the dynamics of the magnetic counterpart from above
despite their different symmetries. In fact, this is the first time, to the best of our knowledge, that propulsion of \emph{truly achiral objects} has been demonstrated.

The electrically polarizable V-micropropeller in Fig.~5(b) is clearly both $\widehat{P}$- and  $\widehat{C}\widehat{P}$-even. General symmetry arguments then imply only that every positive propulsion solution has a complementary solution of opposite propulsion. For the V-shape with permanent magnetic dipole moment [see Fig.~\ref{fig:parity_scheme}(a)] it was shown further that such high symmetry excludes propulsion even in individual instances. This stronger result, however, relied on the explicit form of the equations of motion rather than just symmetry. The equations governing the dynamics of the polarizable electric V-shaped object are, in fact, similar to those governing the $\widehat{C}\widehat{P}$-even ($\widehat{P}$-odd) magnetic propeller [see Fig.~1(b)], apart from the extra scalar factor $\bm{n}\cdot \bm{E}$. \\ \\

\section{Concluding remarks}

To conclude, the shape of an object together with its dipole moment determines its symmetry.
The dipole moment affixed to geometrically achiral planar shape can render it chiral.
Such chiral objects with \emph{intrinsically} broken symmetry can exhibit
steady unidirectional propulsion, resembling that of a helix, when actuated by a rotating field.
In general, however, there could be two distinct in-sync rotational solutions possessing different propulsion velocities.
For certain highly symmetric objects (e.g., $\widehat{P}$- and $\widehat{C}\widehat{P}$-even polarizable propellers) these two velocities average to zero,
while individual structures can propel efficiently due to \emph{spontaneous} symmetry breaking. The theoretical predictions are confirmed experimentally using macro- and microscopic planar magnetized V-shaped propellers driven by a rotating magnetic field.

Our symmetry considerations can be extended to other shapes and actuation schemes. For example, it is interesting to note how these results change upon adding a constant magnetic field along the $z$-axis that breaks the $\widehat{R}_z\widehat{C}$-symmetry (ii), but preserves $\widehat{P}$-symmetry (i). In such a case $\widehat{C}\widehat{P}$-even objects can acquire a nonzero ensemble average velocity. For instance, the V-shape swimmer with dipole moment oriented along the $\bm{e}_1$ symmetry axis ($\widehat{P}$-odd) can, in a certain frequency range, exhibit unidirectional propulsion. Recall that in a planar rotational magnetic field magnetization along any principal axis yielded $\bm{U}={\bm 0}$. The $\widehat{P}$-even object with permanent dipole $\bm{m}$ along a principal axis can still be shown not to propel even when actuated by this less symmetrical field. Dynamics driven by a \emph{conical} rotating magnetic field  (i.e., superposition of a rotating and a constant fields) is, however, beyond the scope of the present paper and will be considered elsewhere.

\section*{Acknowledgement}
This work was supported in part by the German-Israeli Foundation (GIF) via the grant no. I-1255-303.10/2014 (A.M.L. and P.F.), by the Israel Science Foundation (ISF) via the grant No. 1744/17 (A.M.L.), by the Israel Ministry for Immigrant Absorption (K.I.M.) and by the Deutsche Forschungsgemeinschaft (DFG) as part of the project FI 1966/1 (P.F). J.S. and P.F.  thank Donglei Fan and Jianhe Guo for helpful comments regarding the electro-rotation experiments.

\begin{appendices}

\section{Driven propulsion in a rotating magnetic field \label{A}}
\setcounter{equation}{0}
\renewcommand{\theequation}{A\arabic{equation}}

\begin{figure}[b]
\includegraphics[scale=0.7]{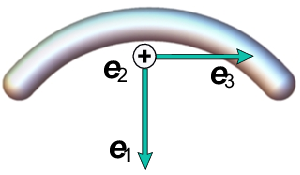}
\caption{Geometry and principal rotational axes of the arc or any planar V-shaped object \cite{note1}.}
\label{fig:arc}
\end{figure}

We consider the propulsion of an arbitrary magnetic object driven by an external uniform magnetic field
${\bm H}=H(\hat{\bm x}\cos \omega t+\hat{\bm y}\sin \omega t)$.
At low Reynolds number an object's translational and rotational velocities $\bm{U}$ and $\bm{\mathit{\Omega}}$
depend linearly on the external torque $\bm{L}=\bm{m}\times\bm{H}$ exerted on it (as the external force vanishes).
\begin{equation}
\bm{U}=\bcalG \cdot \bm{L}\,,\qquad \bm{\mathit{\Omega}}={\bcalF}\cdot \bm{L}\,,
\end{equation}
Here ${\bcalG}$ and ${\bcalF}$ are the coupling and rotation viscous mobility tensors, respectively.
The triad of unit eigenvectors,  $\left\{{\bm e}_1, {\bm e}_2, {\bm e}_3\right\}$ of ${\bcalF}$
makes up the body-frame principal rotation axes. We fix their order such that the corresponding eigenvalues satisfy
${\cal F}_{1}\le{\cal F}_{2}\le{\cal F}_{3}$. For an arc (or any planar symmetric V-shape) the principal rotation axes are shown schematically in Fig.~\ref{fig:arc}.
The lab coordinates unit vectors $\{\hat{\bm x},\hat{\bm y},\hat{\bm z}\}$ are related to
the body frame axes by a rotation matrix ${\cal R}(\varphi,\theta,\psi)$ parameterized
through the three Euler angles $\varphi,\,\theta$ and $\psi$. These Euler angles thus
describe the instantaneous orientation of the object in the lab frame.
The object's magnetization in the body-frame is characterized by a polar angle
$\Phi$ and an azimuthal angle $\alpha$, such that the magnetic moment affixed to the body is given by
${\bm m}=m (s_\Phi c_\alpha {\bm e}_1 + s_\Phi s_\alpha {\bm e}_2 + c_\Phi {\bm e}_3)$,
where we used the compact notation $c_{\Phi}=\cos{\Phi}$, $s_{\alpha}=\sin{\alpha}$, etc.

We are interested in the solutions in which object turns synchronously with the magnetic field, i.e., rotating about the $z$-axis with angular velocity $\bm{\mathit{\Omega}}={\omega}\hat{z}$. This condition reduces the $\bm{\mathit{\Omega}}={\bcalF}\cdot \bm{L}$ into the system of nonlinear equation for the three angles $\psi$, $\theta$ and $\tilde{\varphi}=\varphi-\omega t$.
One can show that this system can further be reduced to a fourth order polynomial equation in $\sin^2\psi$ \cite{prf17}. There can be up to eight in-sync solutions $0\leq\tilde{\varphi},\psi<2\pi,0\leq\theta\leq\pi$ to such an equation. Physically however only stable solution are of real relevance.
The number of stable solutions is found to be at most two.

By symmetry the average velocity in an in-sync solution is in the $z$-direction.
It can be written in a compact covariant form as \cite{prf17}
\begin{equation}
\frac{U_z}{\omega\ell}=\widehat{\bm{\mathit{\Omega}}}\cdot\bCh\cdot\widehat{\bm{\mathit{\Omega}}}\;,\label{eq:Uz}
\end{equation}
where $\bCh$ is a dimensionless \emph{chirality matrix} given by the symmetric part of $\frac{1}{\ell}\, \bcalG \cdot \bcalF^{-1}$ with $\ell$ being the
characteristic length. It is most convenient to write the RHS of Eq.~\ref{eq:Uz} in terms of the body frame components, where $\bCh$ is fixed and $\bm{\mathit{\Omega}}$ expressed using the Euler angles as
\[
\widehat{\bm{\mathit{\Omega}}}=\bm{\mathit{\Omega}}/\omega=\hat{\bm z}=s_{\theta}s_{\psi}{\bm e}_1+ s_{\theta}c_{\psi}{\bm e}_2+ c_{\theta}{\bm e}_3\,.
\]
Notice that the Euler angle $\theta$ defines the precession/wobbling angle between the easy-axis
${\bm e}_3$ and the $z$-axis of the field rotation. One can show that $\bCh$ (in contrast to $\bcalG$)
is independent of the choice of coordinate origin. Under rotation of coordinate frame it transforms
as a (symmetric) pseudo-tensor.

\section{Approximate theory for the magnetic V-shape \label{B}}
\setcounter{equation}{0}
\renewcommand{\theequation}{B\arabic{equation}}

For a V-shaped object (or, in fact, any object possessing two mutually orthogonal planes of reflection symmetry, e.g., the arc in Fig.~\ref{fig:arc}) the only nonzero component of the chirality matrix $\bCh$ in Eq.~(\ref{eq:Uz}) is
$\widetilde{\mathrm {Ch}} \equiv \mathrm {Ch}_{23}=\mathrm{Ch}_{32}=\textstyle \frac{1}{2} \mathcal{G}_{23}({{\cal F}_{2}}^{-1}+{{\cal F}_{3}}^{-1})/a$
with $2a=2\ell$ being the width of the shape and ${\cal F}_{2},{\cal F}_{3}$ the two major eigenvalues of $\bcalF$.
Eq.~(\ref{eq:Uz}) then reduces to
\begin{equation}
\frac{U_z}{\omega a}=\widetilde{\mathrm {Ch}} c_{\psi}s_{2\theta}\;. \label{eq:Uz1}
\end{equation}

The Euler angles $\theta$, $\psi$ and $\tilde{\varphi}$ can be found in a simple closed form assuming cylinder-like rotational anisotropy of the object, i.e.,
${\cal F}_{1} \approx {\cal F}_{2}$, which is an accurate approximation for an arbitrary object (see \cite{scirob}). In such a case, the above mentioned fourth order polynomial equation in $\sin^2{\psi}$ reduces to a quadratic equation.
A cylinder-like object has two distinct well-defined regimes of in-sync
rotation -- tumbling, whereas the easy-axis $\bm e_3$ rotates in the $xy$-plane of the field, and wobbling, whereas $\bm e_3$ undergoes precession with respect to field rotation $z$-axis). The explicit form of the tumbling at low frequencies, $0<{\omega}<\omega_{\mathrm {t\mbox{-}w}}$, is \cite{ML14a}
\begin{equation}
\theta=\pi/2\,,\,\,\psi=-\alpha\,,\,\,{\widetilde{\varphi}}=-\Phi+\textstyle \arccos{\left(\frac{\omega}{\omega_{\mathrm {t\mbox{-}w}}}\right)}\,. \label{eq:t1}
\end{equation}
The  frequency $\omega_{\mathrm {t\mbox{-}w}}$ of the tumbling-to-wobbling transition is $\omega_{\mathrm {t\mbox{-}w}}=\omega_0\cos {\Phi}$, where
$\omega_0$ is the characteristic frequency, $\omega_0=mH {\cal F}_{\perp}$, with ${\cal F}_{\perp}$ being the harmonic mean of the minor rotational mobilities,
 ${\cal F}_{\perp}^{-1} = ({\cal F}_1^{-1}+{\cal F}_2^{-1})/2$.
At higher frequencies, $\omega_{\mathrm {t\mbox{-}w}}<{\omega}< \omega_{\mathrm {s\mbox{-}o}}$, up to the step-out frequency $\omega_{\mathrm {s\mbox{-}o}}=\sqrt{\omega_{\mathrm {t\mbox{-}w}}^2+\omega_0^2 p^2\sin^2 {\Phi} }$ with $p={\cal F}_{3}/{\cal F}_{\perp}$,
there are two complementary wobbling rotational states \cite{prf17}:
\begin{eqnarray}
&\theta_1=\arcsin{\left(\frac{\omega_{\mathrm {t\mbox{-}w}}}{{\omega}}\right)},\,\psi_1=-\alpha-\arcsin{\left(\frac{{{\omega}} c_{\theta_1}}{\omega_0 p s_\Phi}\right)}\,, &\label{eq:w1}\\
&\theta_2=\pi -\theta_1,\,\,\psi_2=-2\alpha-\psi_1\,, & \label{eq:w2}
\end{eqnarray}
and $\widetilde{\varphi}_1=\widetilde{\varphi}_2=0$.
These dual solutions reveal the \emph{bistable} character of the rotational problem usually leading to different propulsion gaits of the object depending on the initial orientation. The substitution of Eqs.~(\ref{eq:w1}), (\ref{eq:w2})  into eq.~(\ref{eq:Uz1}) gives the closed-form expression for the in-sync propulsion velocity:
\begin{eqnarray}
\frac {U_z}{\omega a}&=& \frac {2\widetilde{\mathrm {Ch}}}{\gamma}\, \left[ -\sin \alpha \left( 1-\frac {\omega_{\mathrm {t\mbox{-}w}}^2}{\omega^2}\right)
\pm \right. \nonumber \\
&& \quad \left.\cos \alpha \sqrt{1-\frac {\omega_{\mathrm {t\mbox{-}w}}^2}{\omega^2}} \sqrt{\frac {\omega_{\mathrm {s\mbox{-}o}}^2}{\omega^2}-1}
 \right]\,, \label{eq:U2}
\end{eqnarray}
where $\gamma=p \tan \Phi$.

Thus, Eq.~(\ref{eq:U2}) gives the propulsion velocity of V-shaped magnetic object in the
high-frequency domain, $\omega_{\mathrm {t\mbox{-}w}}<{\omega}< \omega_{\mathrm {s\mbox{-}o}}$,
assuming cylinder-like rotational anisotropy. In the low-frequency domain, $0<{\omega}<\omega_{\mathrm {t\mbox{-}w}}$,
the propeller tumbles without translation. At frequencies above the step-out,
$\omega>\omega_{\mathrm {s\mbox{-}o}}$, the object can no longer rotate in-sync with
the field and the transition to the asynchronous regime takes place. During asynchronous
rotation the propulsion velocity rapidly diminishes with increasing
frequency. This regime is beyond the scope of the present paper.
Depending on the orientation of an objects magnetization there are two distinct propulsion
scenarios considered in detail in the main text. The first case corresponds to out-of-plane
[OOP, e.g. Fig.~1(c) of the paper] magnetization, with $\cos{\alpha}=0$ and the second case to
in-plane [IP, e.g. Fig.~1(b) of the paper] magnetization, with $\sin{\alpha}=0$.
Below we consider both these cases in detail.

\subsubsection*{Out-of-plane magnetization}
When $\cos \alpha=0$, there are two non-trivial components of the propeller magnetization
${\bm m}$ -- along the directions ${\bm e}_2$ and ${\bm e}_3$ as shown in Fig.~1(c) in the main text.
In this case the dual rotational states results in the same propulsion gait with the velocity
\begin{equation}
\frac{U_z}{\omega a}= -\frac {2\widetilde{\mathrm {Ch}}}{\gamma}\,
\sin \alpha \left( 1-\frac {\omega_{\mathrm {t\mbox{-}w}}^2}{\omega^2}\right)
\,. \label{eq:OOP}
\end{equation}
Thus, for positive values of the parameter $\widetilde{\mathrm {Ch}}>0$ the V-shape object
with the magnetic component in the direction of ${\bm e}_2$ [$\alpha=\pi/2$ as in
Fig.~1(c)] moves in the $-z$ direction ($U_z<0$), i.e., similar to a left-handed helix. In contrast,
the V-shaped object with the magnetic component pointing in the $-{\bm e}_2$ direction
($\alpha=-\pi/2$),  moves along the $z$-axis ($U_z>0$) similarly to a right-handed helix. Due to the freedom in the choice of
orientation of the principal axes of rotation $\bm{e}_i$ we assume here and thereafter w.l.o.g. that $0<\Phi<\pi/2$ and $\gamma=p\tan{\Phi}>0$.

We emphasize again that this helix-like propulsion corresponds to both
rotational branches of the solution, e.g, the right-handed object under
clock-wise rotation (CW) of the external field can move in the $+z$ direction by two different ways:
the object easy-axis ${\bm e}_3$ may acquire either an obtuse or acute angle with the field rotation
$z$-axis. In the scaled-up experiment the cm-sized arc's two halves were painted in two different
colors -- red and green. In case of an OOP magnetization it was shown to propel with either the
red or the green end forward.

\subsubsection*{In-plane magnetization}
When $\sin \alpha=0$, there are two in-plane magnetization components ${\bm m}$ -- along the
${\bm e}_1$ and ${\bm e}_3$ axes as shown in Fig.~1(b) in the main text. In this case the dual
in-sync rotational states yield propulsion in opposite directions:
\begin{equation}
\frac {U_z}{\omega a}= \pm \frac {2\widetilde{\mathrm {Ch}}}{\gamma} \cos \alpha \sqrt{1-\frac {\omega_{\mathrm {t\mbox{-}w}}^2}{\omega^2}}
\sqrt{\frac {\omega_{\mathrm {s\mbox{-}o}}^2}{\omega^2}-1}\,. \label{eq:IP}
\end{equation}
This means that the two symmetric (in polar angle $\theta$) solution branches
(\ref{eq:w1}) and (\ref{eq:w2}) result in propulsion
with the same speed in opposite directions. In contrast to the OOP-magnetized V-shape,
the IP-magnetized arc can propel either along the $z$-axis of the field rotation
or anti-parallel to it. Moreover, unlike the OOP-magnetization, the IP-magnetized V-shape
always propels while keeping the same
orientation (of ${\bm e}_3$) with respect to the field, meaning that
depending on the value of the azimuthal magnetization angle $\alpha=0$
(or, respectively, $\alpha=\pi$) it will move either with the red
(or, respectively, the green) end forward regardless of the sense of the fields
rotation (CW or CCW).

\section{Parameter estimates for the magnetic arc \label{C}}
\setcounter{equation}{0}
\renewcommand{\theequation}{C\arabic{equation}}

In the experiments we used the cm-size propeller [see Fig.~2(b) of the paper] in the shape of a circular arc with the centerline radius $9.56$~mm,
waist $2a=2$~mm and central angle of $119^\circ$. The mobility tensors were computed numerically using the particle-based multipole expansion method (see, e.g., \cite{Filippov}
for a detailed description). We found
\begin{eqnarray}
&& {\bcalF}=\frac {10^{-4}}{\eta a^3}\mathrm {diag}(2.36,2.42,14.95)\,, \nonumber \\ 
&& \mathcal{G}_{23}=\frac {2.14\cdot10^{-4}}{\eta a^2}\,, \label{eq:fg}
\end{eqnarray}
where $\eta$ is the fluid dynamic viscosity. The value of the pseudo-chirality coefficient is then
$\widetilde{\mathrm {Ch}} \equiv \textstyle \frac{1}{2} \mathcal{G}_{23}({{\cal F}_{2}}^{-1}+{{\cal F}_{3}}^{-1})/a=0.515$.
The transverse anisotropy parameter proves to be quite small, $\varepsilon=({\cal F}_2-{\cal F}_1)/({\cal F}_2+{\cal F}_1)=0.011$, suggesting
that the above cylindrical approximation is in fact quite accurate.  The longitudinal anisotropy parameter is $p={\cal F}_{3}/{\cal F}_{\perp}=6.25$ and the the transverse rotational mobility ${\cal F}_{\perp}=2/({\cal F}_1^{-1}+{\cal F}_2^{-1})=2.39\cdot 10^{-4}/(\eta a^3)$. Using the values of $H=300$~Oe, $m=0.975$~emu, $\eta=10$~P for glycerol,
and $a=0.1$~cm we can estimate the characteristic frequency as $\omega_0=m H {\cal F}_{\perp} \approx 7$~s$^{-1}$.

\subsubsection*{Out-of-plane magnetization, Fig.~3(a) of the paper}
The  polar magnetization angle was $\Phi=15^{\circ}$ and the azimuthal angle $\alpha=-90^{\circ}$
corresponding to a right-handed helix, resulting in $\gamma=p \tan \Phi=1.67$.
The estimated frequencies of the tumbling-to-wobbling transition and the step-out are
$\omega_{\mathrm {t\mbox{-}w}}=6.65$~s$^{-1}$ and $\omega_{\mathrm {s\mbox{-}o}}=13.22$~s$^{-1}$,
respectively, giving $\nu_{\mathrm {t\mbox{-}w}}=\omega_{\mathrm {t\mbox{-}w}}/2\pi=1.06$~Hz and
$\nu_{\mathrm {s\mbox{-}o}}=\omega_{\mathrm {s\mbox{-}o}}/2\pi=2.10$~Hz. Substitution of the values
into Eq.~(\ref{eq:OOP}) gives the propulsion velocity as a function of actuation frequency $\nu$
for the case of OOP magnetization of the V-shaped object:
\begin{equation}
\frac{U_z}{\omega a}= 0.614 \left( 1-\frac {1.06^2}{\nu^2}\right)\,. \label{eq:OOP2}
\end{equation}

The experimental data points ($\circ$, $\bullet$) in Fig.~3a of the paper are in excellent
agreement with the theoretical prediction in Eq.~(\ref{eq:OOP2}) for frequencies $\nu<1.6$~Hz.
Notice that stable in-sync rotation in the frequency interval $1.6$~Hz$<\nu<2.1$~Hz
does not materialize in the experiments. This is due to the fact that for frequencies $\nu>1.6$~Hz
[dashed line in Fig.~3(a)] the basin of attraction in terms of initial orientations, of the in-sync solution,
shrinks rapidly in size, while for $1.06$~Hz~$<\nu<1.6$~Hz
[solid line in Fig.~3(a)] the solution converges to the steady-state for an arbitrary initial
orientation (i.e., basin of attraction filling all space).
We could not observe the stable solutions at $1.6$~Hz~$<\nu<2.1$~Hz,
as it was not possible to smoothly vary the actuation frequency on-the-fly in the experiments.
Due to the finite size of the cuvette we started each experiment from rest with the arc placed
roughly at the center of the cuvette and the rotating field was switched on abruptly with
a particular (fixed) frequency $\nu$.  At $\nu>1.6$~Hz, the aperiodic tumbling of the V-shape takes
place ($\vartriangle$, $\blacktriangle$) in a complete agreement with the theoretical prediction
(dotted line) based on numerical integration of the dynamical system of Eqs.(5)--(7) in \cite{prf17},
for the Euler angles using the above estimated parameters and the geometry of the arc.

\subsubsection*{In-plane magnetization, Fig.~3(b) of the paper}
The  polar angle of magnetization in this case was $\Phi=13^{\circ}$ and the azimuthal angle was close to $\alpha=180^{\circ}$. The estimated parameters are $\gamma=p \tan \Phi=1.33$,  $\nu_{\mathrm {t\mbox{-}w}}=1.10$~Hz and $\nu_{\mathrm {s\mbox{-}o}}=1.80$~Hz. Substituting these values into Eq.~(\ref{eq:IP}) gives the propulsion velocity of the V-shape object as a function of actuation frequency $\nu$ for the IP-magnetization:
\begin{equation}
\frac{U_z}{\omega a}= \pm 0.775 \sqrt{ 1-\frac {1.10^2}{\nu^2}}\sqrt{ \frac {1.80^2}{\nu^2}-1}\,. \label{eq:IP2}
\end{equation}

The Eq.~(\ref{eq:IP2}) represents two symmetric (with respect to the $\nu$-axis) branches, however,
in reality, perfect IP magnetization is hard to achieve. Best fits of the theory to the experiments [see $\circ$, $\bullet$ in Fig.~3(b)]
suggest minor imperfection in the azimuthal magnetization angle, $\alpha\approx 183^{\circ}$, resulting in a slight asymmetry
between the two solution branches in (\ref{eq:IP2}) shown in Fig.~3(b) by the continuous (blue and red) lines.
The agreement between the theoretical prediction (solid colored lines) and the experimental results is very good. As anticipated by the theory,
both branches are equally accessible upon CW and CCW actuation. As for the case of OOP-magnetization and for the same reason, the high-frequency stable in-sync solutions (dashed colored lines) is hard to realize in the experiments. Therefore aperiodic tumbling ($\vartriangle$, $\blacktriangle$) takes over before the step-out at $\nu<1.8$~Hz is attained.
And as for the OOP magnetization, the theoretical prediction of the aperiodic tumbling based on numerical integration of the full dynamical system in \cite{prf17} (dotted line) shows excellent agreement with the experimental results. Notice that the measurements were taken over a whole day during which the temperature and humidity raised resulting in somewhat lower (about 40\%) viscosity of the glycerol as the viscosity of the pure glycerol is quite sensitive to temperature and humidity. This change resulted in proportional increase of the characteristic (dimensional) tumbling-to-wobbling and step-out frequencies. Therefore, the values of frequencies of the afternoon measurements were down-scaled by a constant factor of $1.4$ to match the measurements taken in the morning hours.

\section{Dual solution, in-plane magnetized arc \label{D}}

\begin{figure}[tb]
\includegraphics[scale=.9]{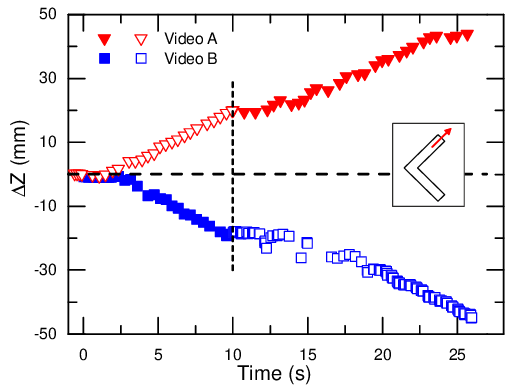}
\caption{\label{fig:switching} Displacement of the magnetic arc depicted in the inset vs. time. Two experiments were conducted at the actuation frequency $1.25$~Hz  (triangles and squares). Open symbols stand for CCW, whereas full symbols denote CW rotation of the external field. In each case the field rotation sense is abruptly reversed after 10~s (vertical dashed line) resulting (after a short transient) in the arc's propulsion along the same direction with the same speed as before.}
\end{figure}

The displacement in the $\pm z$ direction for the IP-magnetized arc
[as in Fig.~1(b) in the main text], is shown in Fig.~\ref{fig:switching} over time.
The data was extracted from Video~\#2 \cite{SM}, in which the direction of the field rotation was suddenly
reversed ($\bm{\mathit{\Omega}}\rightarrow -\bm{\mathit{\Omega}}$)
after the arc started to move with a constant steady velocity
$U_z$. Upon this reversal the arc tumbles for a few revolutions, before acquiring a constant velocity
equal to $U_z$ in the same direction as before.
Recalling that our $\hat{\bm z}$ direction was defined by $\bm{\mathit{\Omega}}$
this is, in fact, equivalent to a change of the sign
of $U_z={\bm U}\cdot\hat{\bm z}$, whereas its absolute value is the same before and after the
field reversal in both experiments ($|U_{z}|\approx 2.4$~mm/s). These field reversal experiments
confirm unequivocally the theoretical prediction in Eq.~(\ref{eq:IP}) that the propulsion direction
of IP-magnetized arc is independent of the field rotation sense.

\section{Achiral magnetic arc \label{E}}

\begin{figure}[tb]
\includegraphics[width=0.45\textwidth]{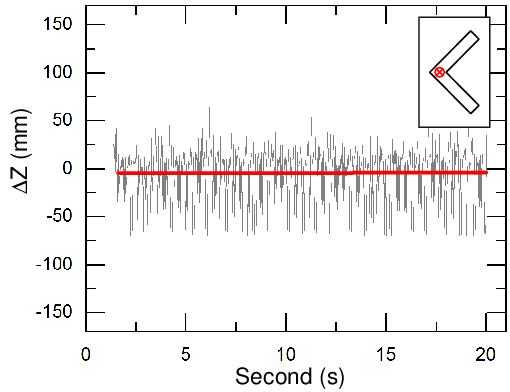}\hfill
\includegraphics[width=0.45\textwidth]{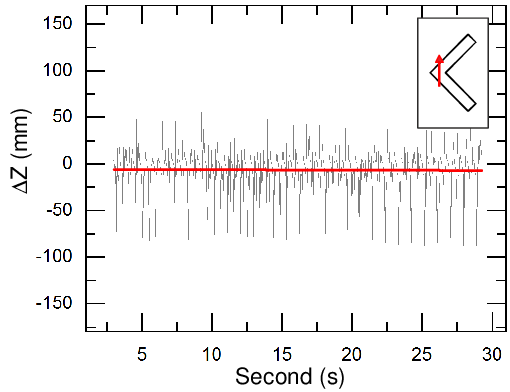}
\caption{\label{fig:No_prop} Displacement plots of two achiral cm-sized arcs, each schematically depicted in the corresponding inset. The actuation frequency for the left one is 5.25\thinspace Hz while the right one was rotated with 1.05\thinspace Hz. The slope of the fits is close to zero for both cases, indicating that arcs with these symmetries are non-propulsive.}
\end{figure}

In Fig.~\ref{fig:No_prop} the displacement of the magnetic arc is plotted vs. time for two shapes shown in the respective insets. The data was extracted from Video~\#3 \cite{SM}. In both cases the magnetic moment is oriented along one of the principal axes of rotation rendering the arcs $\widehat{P}$- and $\widehat{C}\widehat{P}$-even and therefore truly achiral. Thus, their dynamics is precession-free resulting in no net propulsion. This is readily seen from the negligible slope of the linear fit (red lines). Notice that the step-out frequencies for the two arcs shown in Fig.~\ref{fig:No_prop} are different. The arc shown in the right panel has a step-out frequency of $\nu_{\mathrm {s\mbox{-}o}}\approx 1.2$~Hz compared to the one shown in the left panel with  $\nu_{\mathrm {s\mbox{-}o}}\gtrsim 5$~Hz.

\section{$\widehat{P}$-odd, $\widehat{C}\widehat{P}$-even magnetic V-shaped microcolloids \label{F}}

\begin{figure}[tb]
\includegraphics[scale=0.90]{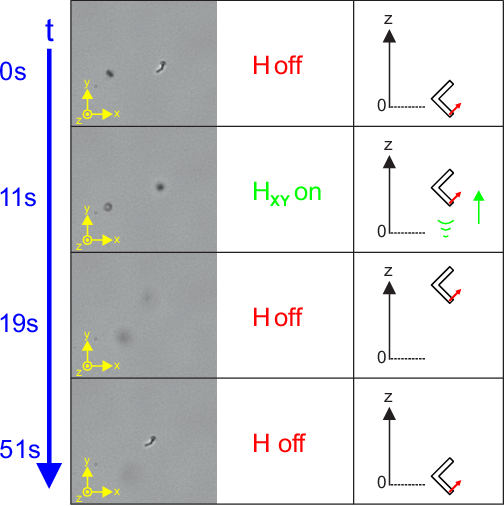}
\caption{\label{fig:GLAD_z} Time evolution of a magnetized V-shaped microcolloid moving along the $z$-axis upon a rotating field in the $xy$-plane, i.e. it moves out of the focal plane of the microscope. In the right column the $z$-position of the structure is schematically depicted.}
\end{figure}

The Video~\#4 \cite{SM} shows the ability to control the trajectory of the  ($\widehat{C}\widehat{P}$-even, $\widehat{P}$-odd) microcolloids by switching the external field rotation plane between all three principal planes. Notice that the $z$-axis is now perpendicular to the microscope focal plane in contrast to the microcolloid experiments described in the main text (see Fig.~4), whereas the $z$-axis belonged to the focal plane of the microscope. The frames in Fig.~\ref{fig:GLAD_z} correspond to a movement out-of-plane and were extracted at the specified times.  They show the ability of $\widehat{C}\widehat{P}$-even V-shaped microsctructures can propel against gravity. One can see that the structure is in the microscopes' focus at the beginning of the video and starts to move out-of-focus when the magnetic field is turned on in the $xy$-plane at 11~s. The  magnetic field is then switched off at 19~s and the structure sediments and reappears in the focal plane of the microscope. Waiting even longer leads to further sedimentation and the image de-focuses again (not shown). This indicates that the trajectory of this microcolloid can be controlled in every spatial direction, including $x, y$ and $z$.

\section{Achiral ($\widehat{P}$-, $\widehat{C}\widehat{P}$-even) electrically polarizable V-shaped microcolloids \label{G}}
\begin{figure}[tbh]
\includegraphics[scale=0.90]{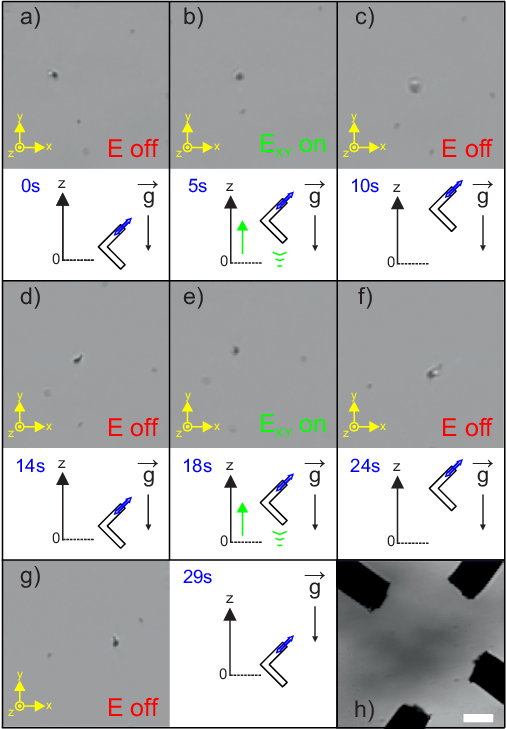}
\caption{\label{fig:extendedEarc} Extended time-evolution of the achiral V-shaped microcolloid driven by an electrical field, including video frames and a schematic visualization (a-g). The double-headed arrow visualizes the induced magnetic moment by the external electric field. Image of the experimental four-electrode setup (scale bar length $100$~$\mu$m) (h).}
\end{figure}

The same propulsion behavior as described in the previous section was demonstrated for a microcolloid driven by an electric instead of a magnetic field. This is shown in Video~\#6 \cite{SM}, from which the frames shown in Fig.~\ref{fig:extendedEarc} were extracted; the $z$-axis is now perpendicular to the focal plane of the microscope. The applied electric field strength was around $E=7.5$~kV/m with an AC frequency of $500$~kHz, while the V-shaped microstructures rotate with a few Hz as seen in the Video. It is obvious that the structure is initially at rest. As the AC electric fields starts to rotate in the $xy$-plane, the structure moves out-of-focus (along $+z$-direction). After switching off the electric field the structure sediments due to gravity. This sequence was repeated twice in a row. The behavior of the electric dipolar particle thus resembles the propulsion dynamics of the magnetic V-shaped microcolloid in Fig.~\ref{fig:GLAD_z}, but this time with a truly achiral (i.e., $\widehat{P}$- and $\widehat{C}\widehat{P}$-even) object. Notice that the magnetic analogue of such an achiral propeller in Fig. 1a in the main text does not propel. The electro-rotation of the structures is highly dependent on the geometry of the structure. Note that the visible drift in the $xy$-plane is most likely due to a small gradient of the electric field, as the four-electrode setup shows some asymmetry [see Fig.~\ref{fig:extendedEarc}(h)].

\end{appendices}


\begin{thebibliography}{99}
\bibitem{time_rev} The possibility to classify solutions into $\widehat{T}$-dual pairs related by
$(\psi,\theta,\varphi)\leftrightarrow(\psi+\pi,\pi-\theta,-\varphi)$ may however be convenient sometimes..
\bibitem{phase}
Alternatively $\widehat{R}_z$ can be replaced by a $\pi$-phase shift of the field, $t\mapsto t+\pi/\omega$.
\bibitem{note1} Notice that depending on slenderness of the V-shaped object the minor principal axes may be interchanged,
${\bm e}_1 \leftrightarrows {\bm e}_2$ \cite{prf17}.
\end{thebibliography}
\end{document}